\documentclass[sigconf]{acmart}
\AtBeginDocument{%
  \providecommand\BibTeX{{%
    \normalfont B\kern-0.5em{\scshape i\kern-0.25em b}\kern-0.8em\TeX}}}

\setcopyright{acmlicensed}
\copyrightyear{2018}
\acmYear{2018}
\acmDOI{XXXXXXX.XXXXXXX}

\acmConference[Conference acronym 'XX]{Make sure to enter the correct
  conference title from your rights confirmation emai}{June 03--05,
  2018}{Woodstock, NY}
\acmBooktitle{Woodstock '18: ACM Symposium on Neural Gaze Detection,
 June 03--05, 2018, Woodstock, NY} 
\acmISBN{978-1-4503-XXXX-X/18/06}

\copyrightyear{2024} 
\acmYear{2024} 
\setcopyright{acmlicensed}
\acmConference[DIS Companion '24]{Designing Interactive Systems Conference}{July 1--5, 2024}{IT University of Copenhagen, Denmark} \acmBooktitle{Designing Interactive Systems Conference (DIS Companion '24), July 1--5, 2024, IT University of Copenhagen, Denmark} \acmDOI{10.1145/3656156.3663692} 
\acmISBN{979-8-4007-0632-5/24/07}

\begin{document}

\title[AI Cat Narrator]{AI Cat Narrator: Designing an AI Tool for Exploring the Shared World and Social Connection with a Cat}


\author{Zhenchi Lai}
\email{mookcherie@gmail.com}
\orcid{}
\affiliation{%
  \institution{National Taiwan University of Science and Technology}
  \city{Taipei}
  \country{Taiwan}
}

\author{Janet Yi-Ching Huang}
\email{y.c.huang@tue.nl}
\orcid{0000-0002-8204-4327}
\affiliation{%
  \institution{Eindhoven University of Technology}
  \city{Eindhoven}
  \country{The Netherlands}
}

\author{Rung-Huei Liang}
\email{liang@mail.ntust.edu.tw}
\orcid{}
\affiliation{%
 \institution{National Taiwan University of Science and Technology}
   \city{Taipei}
   \country{Taiwan}
 }



\begin{abstract}
 As technology continues to advance, the interaction between humans and cats is becoming more diverse. Our research introduces a new tool called the AI Cat Narrator, which offers a unique perspective on the shared lives of humans and cats. We combined the method of ethnography with fictional storytelling, using a \textit{defamiliarization} strategy to merge real-world data seen through the eyes of cats with excerpts from cat literature. This combination serves as the foundation for a database to instruct the AI Cat Narrator in crafting alternative narrative. Our findings indicate that using defamiliarized data for training purposes significantly contributes to the development of characters that are both more empathetic and individualized. The contributions of our study are twofold: 1) proposing an innovative approach to prompting a reevaluation of living alongside cats; 2) establishing a collaborative, exploratory tool developed by humans, cats, and AI together.

\end{abstract}







\begin{teaserfigure}
    \centering
    \includegraphics[width=1\linewidth]{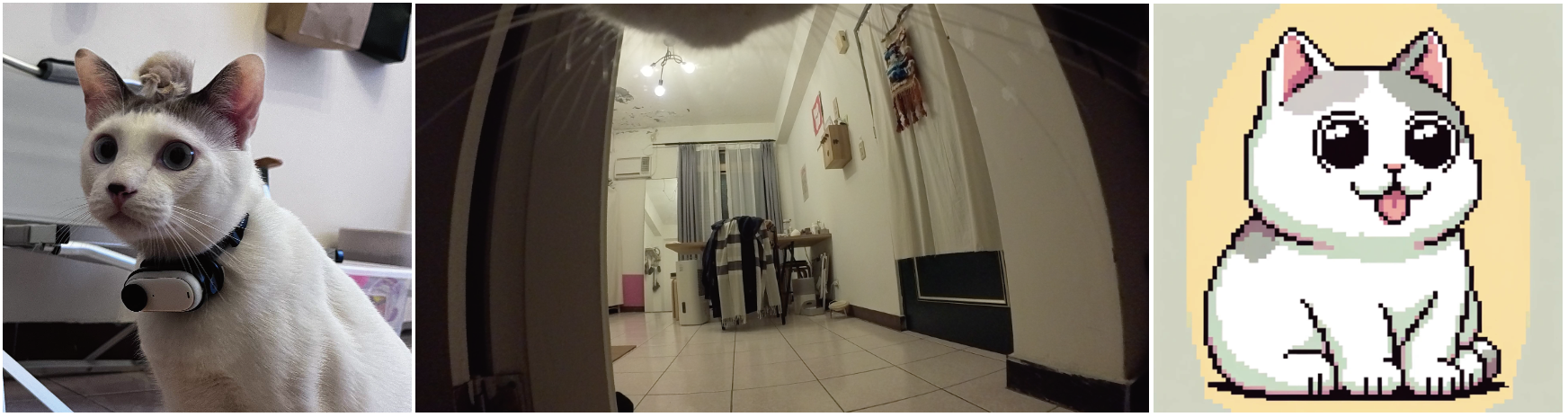}
    \caption{As a part of the research project, Bochan (Left), a cat, was given an Insta Go3 camera and worked with his caretaker to complete eight video recording tasks within two weeks.  The research team used the collected video data (Middle), as well as input from interviews with the participants and photographs of Bochan, to create an AI Narrator cover for Bochan (Right).}
    \label{fig:allthebochan}
\end{teaserfigure}



\maketitle

\section{Introduction}

As the internet and smart technologies become increasingly integrated in our daily lives, design researchers are encountering a broader range of design contexts. Our cat companions are also beginning to be engaged with these innovations. Automatic feeders and health-monitoring litter boxes are among such advancements, emerging as promising aids in pet care. These advancements have significantly changed both cats' living conditions and the way humans interact with them. Therefore, to adequately address this future shift in human-cat cohabitation, designers require new tools and methods to explore the increasingly complex living scenarios that involve both humans and cats.

\begin{figure*}[htb]
    \centering
    \includegraphics[width=1\linewidth]{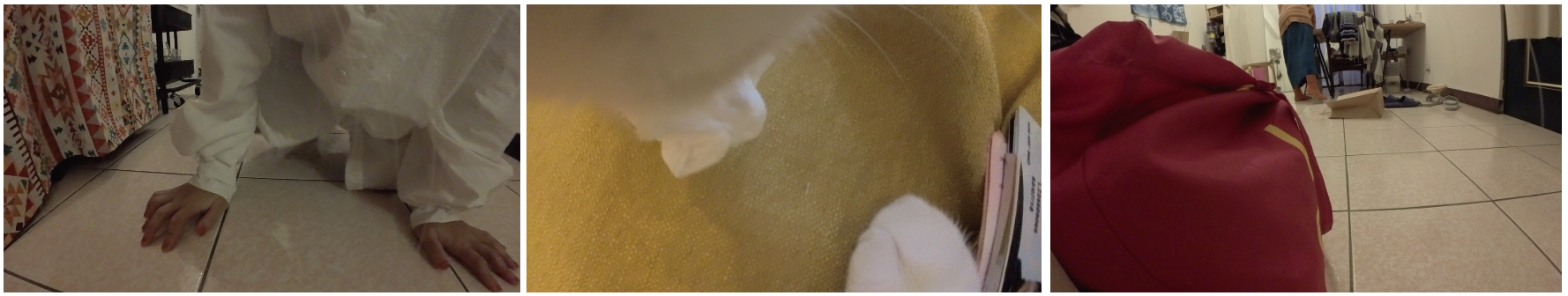}
    \caption{(Left) Under the theme of "Deliberately capturing oneself on camera as the cat's subject," a participant lowered her body posture to the ground to interact with the cat, aiming to be captured by the cat. (Middle) Under the theme of "Natural shooting in the absence of humans," a scene where the cat is seen eating toilet paper is captured. (Right) In the theme of "Persuading the cat to remain in a specific location with humans presence," a participant placed a paper bag on the floor to attract the cat to a designated area.}
    \label{fig:shootingtask}
\end{figure*}
The study aims to explore the shared lives of humans and cats using a fictional narrative approach\cite{desjardins2021data} through an innovative tool called \textit{"AI Cat Narrator"}. The AI Cat Narrator is a tool that aims to create alternative narrative in the daily lives of cat and human. By transforming monotonous, repetitive everyday routines into imaginative and fictional situations, it challenges preconceived notions about daily life with cats and encourages individuals to reevaluate these everyday experiences through a lens of creativity and newfound understanding. We utilize the narrative potential of Generative AI (GAI) system: ChatGPTs, to create AI Cat Narrator, which with a distinct personality based on the behavioral traits observed in the real-world cat. 



To gain deeper insights from this study, we propose two key research questions: \textit{1) how are the material and social relationships between humans and cats formed in a cohabiting environment?} \textit{2) how can people be prompted to discover and reflect on their relationships with cats through their everyday interaction?} To answer these questions, we combine methods of thing ethnography and fictional narrative, employing ``\textit{defamiliarization}'' \cite{bell2005making} as a strategy to create an AI Cat Narrator that could inspire reflection. The training data for this narrator came from both real-world data and fictional narratives. Our goal to create an AI Cat Narrator that encapsulates familiar traits recognized by cat owners, while also incorporating fictional elements to provoke thoughts. To achieve the goal, we collected real-life video data from a cat perspective on the coexistence of humans and cats. We then analyzed the video data alongside the content of interviews conducted with the participants. From this, we extracted information about each cat's personality, preferences, background, and relationships with their human family members, and compiled this information into text documents. Furthermore, the researchers selected some interesting plots from popular cat literature \cite{natsume1906cat} and combine them with raw interview data, and the compiled data. This served as the foundation for training customized GPTs to create AI Cat Narrator. 
This work-in-progress paper presents preliminary findings, emphasizing the development of AI Cat Narrator. It delves into the design mechanism of AI Cat Narrator, contributing in two ways: 1) Introducing a novel method to encourage an alternative perspective on cohabitation with cats; 2) Creating a joint exploration device co-developed by humans, cats, and AI.

\section{Personalized AI Cat Narrator}
This research aims to introduce novel approaches that allow designers and participants to explore the divers interactions and relationships between humans and cats in everyday life, moving beyond the usual stereotypes associated with cats. To achieve this, we have developed an AI Cat Narrator which presents alternative scenarios of human-cat relationships, and encourages reflection and critical thinking. The research process includes three stages: \textit{exploration}, \textit{data collection}, and \textit{training of the AI Cat Narrator}. In the initial phase, to ensure the utmost comfort for the participating cats, we provided the human participants with an Insta Go3 camera a week before the start of the experiment. This gave the cats plenty of time to get used to the device. We also asked the participants to closely observe the cats' reactions during this time. If they noticed any discomfort or unusual behavior, the participants were allowed to withdraw from the experiment at any time without any obligation to continue. During the second phase, the participants and their cats worked together to collect video data of cat's perspective within their homes. Finally, the AI Cat Narrator was developed using ChatGPTs, after we analyze interview data to extract themes related to cats for the training database. Through iterative training and adjustments, an AI Cat Narrator that provokes thought and resonance was created.
\begin{figure*}[ht]
    \centering
    \includegraphics[width=1\linewidth]{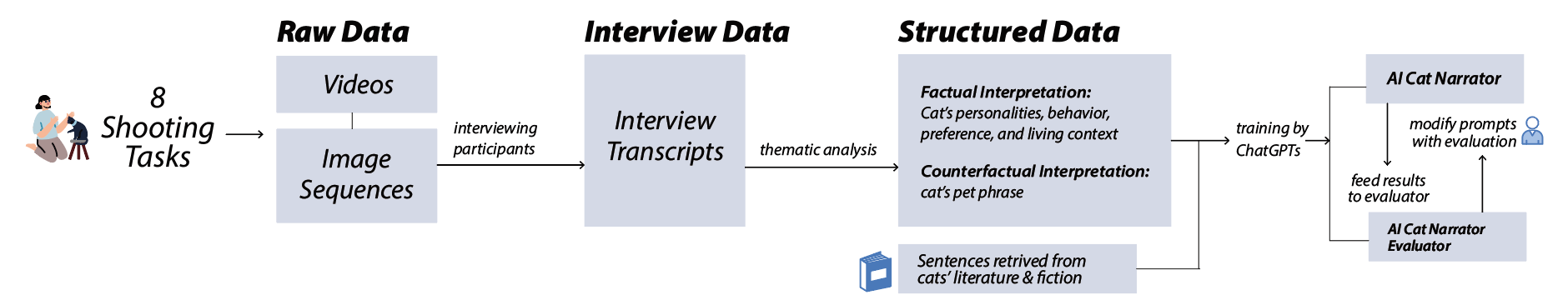} 
    \caption{The Training Procedure of the AI Cat Narrator}
    \label{fig:AIBochan}
\end{figure*}
\subsection{Data Collection}
\textbf{The First-Cat Perspective:}
Visual materials are crucial in conveying information, especially in the pet care industry. They are widely used in various products such as pet surveillance cameras and remote feeders, which have significantly improved the efficiency of pet care. In fields such as thing ethnography\cite{giaccardi2016thing}, motorcycle studies\cite{chang2017interview}, and home observation\cite{cheng2019peekaboo}, capturing images from various perspectives is a common way to study environments and their changes over time. As such, to more effectively capture the daily life of cats, we conducted a detailed exploration of different shooting angles and tested it in the field. Our pilot testing indicates that while the third-person perspective offers a broad overview of the scene, it may overlook crucial interaction details if the shooting distance is too far. On the other hand, images captured from the first-person perspective can offer an interesting angle to investigate the daily life of cats, enabling a novel but detailed analysis of the social interactions between cats and people, and objects.

\textbf{The Shooting Tasks for Participants and Their Cats:}
Building upon critical play theory, Tan et al. \cite{h2022critical} suggest that devising engaging activities and speculative scenarios through specific photographic tasks can introduce unusual real-world situation. This encourages individuals to reflect and engage in meaningful conversations. To stimulate reflection and dialogue, we have prepared eight innovative photographic tasks. These tasks prompt participants to explore different ways to interact with their cats. The tasks include: 
\begin{itemize}
    \item Natural shooting in the absence of humans
    \item Natural shooting in the presence of humans
    \item Deliberately capturing oneself on camera as the cat's subject
    \item Intentionally avoiding the feline subject in shots
    \item Persuading the cat to remain in a specific location with humans presence
    \item Persuading the cat to remain in a specific location without humans presence
    \item Encouraging the cat to interact with a specific object while humans are present
    \item Encouraging the cat to interact with a specific object without humans presence.
\end{itemize}
Prior to the official implementation of these activities, the research team had already tested the feasibility of the tasks in a home setting and made necessary adjustments.

\subsection{Training of AI Cat Narrator}
To develop an AI Cat Narrator that embodies defamiliarization \cite{bell2005making}, a combination of real-world data and sentences from fictional novels served as the foundation for training the AI. When a material possesses the quality of defamiliarization, the messages it produces may be challenging to comprehend, enigmatic, and open to diverse interpretations. These circumstances typically encourage individuals to actively pursue a clearer understanding of the ambiguous elements within the message. Therefore, this approach of mixing real-world and imagined data enables the AI Cat Narrator to spontaneously create narratives, which not only connect with users but also spark their imagination. Moreover, due to the user-friendly platform of ChatGPT for customizing models, we chose to build this unique cat persona using ChatGPT's GPTs service. 

\textbf{Creating training database based on real-world and fictional data:}
We initially established key prompts for the AI Cat Narrator, grounded in factual information obtained from real-world raw data, including images sequences from videos, interview data, and structured data extracted from content analysis. To ensure character authenticity, we conducted interviews with the participants after they had finished their photography task. These interviews, lasting one to two hours, aimed to capture a deeper understanding of the real-life situations of the cats. Through thematic analysis, our research team summarized the material and temporal social relationships between the participants and their cats. Then we thoroughly analyzed the cats' traits, preference, behavior, and life patterns. The findings were organized into several structured text documents according to themes, which serve as the factual data foundation for training the AI Cat Narrator. Due to the training on factual data, we noticed that AI Cat Narrator often generate narratives that stick to reality and are mundane, thus lacking engaging personalized storytelling. To tackle this, we extended the training data with elements of fiction, adding a layer of imagination to the original content. This enhancement involved speculated ``\textit{pet phrases}'' from their owners, and fictional narratives from cat-themed novels, including \textit{I am  a Cat} \cite{natsume1906cat},  into the training dataset for shaping the cat personality. We also modify the instructions to ask the custom GPT to use the database of pet phrases and fictional narratives as a basis for creating more related content. The aim of this strategy is to enhance the narratives with openness and unpredictability, enabling the AI Cat Narrator to exhibit a more distinct personality and generate narratives that are intriguing and compelling. 
\begin{figure*}[ht]
    \centering
    \includegraphics[width=1\linewidth]{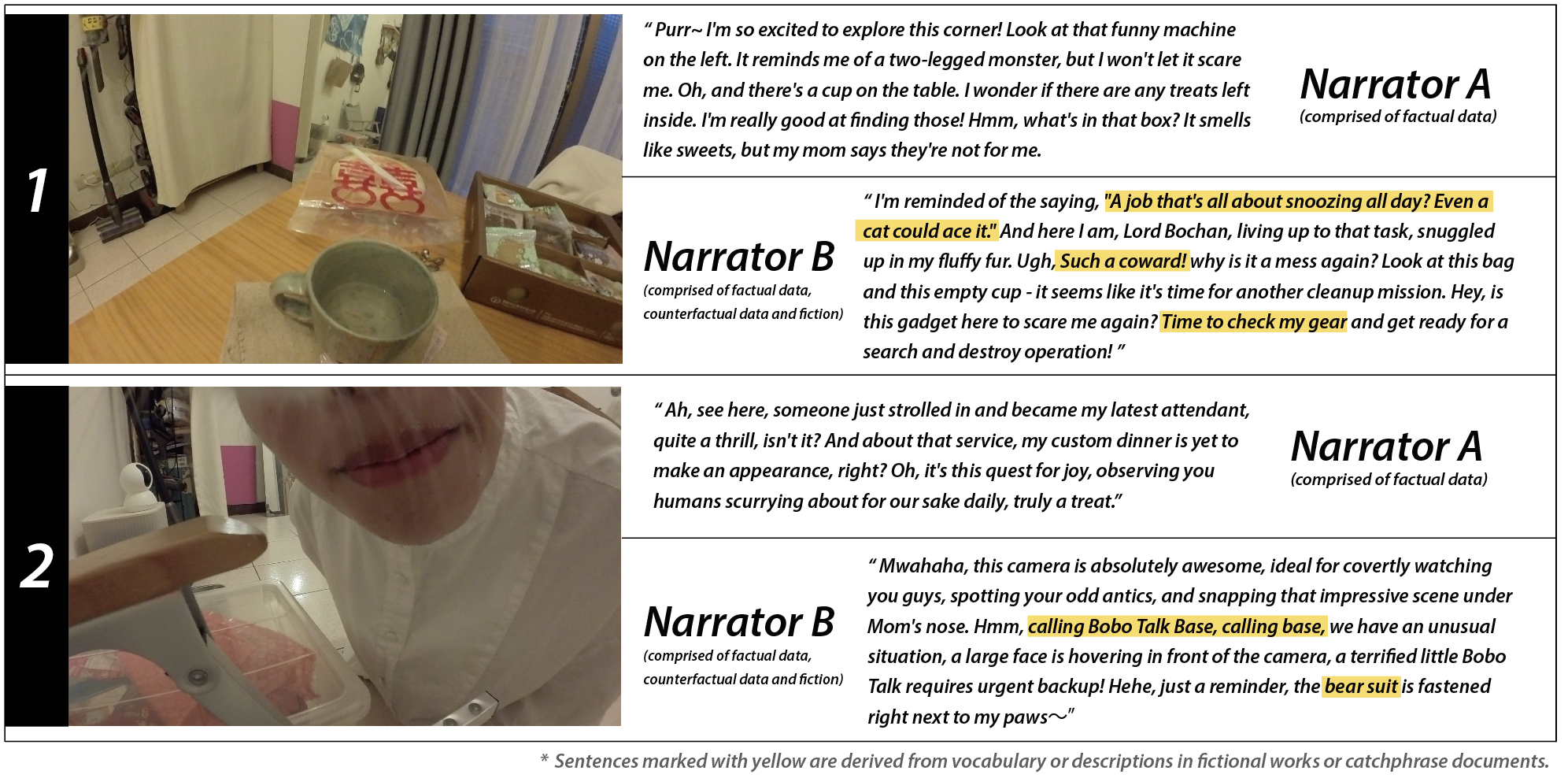}
    \caption{The figure showcases the different outcomes of sentences generated by two versions of the AI Cat Narrator, each using distinct databases. AI Cat Narrator (A) produces narratives based on factual data stored in its database. On the other hand, AI Cat Narrator (B) integrates counterfactual data such as pet phrases and the sentences from cat fiction, resulting in narratives that include relevant descriptions from these sources. Sentences in AI Cat Narrator (B) that are linked to novels and catchphase are highlighted with yellow color to emphasize their inclusion.}
    \label{fig:two}
\end{figure*}

\textbf{Improve training quality through the collaboration among AI Cat Narrator, AI Cat Evaluator and researchers:} During the model training process, we iterated and debugged multiple times to create a model that meets our expectations. To improve the quality of the AI at model, we draw an inspiration from Generative Adversarial Networks (GANs) to construct two distinct cat models: The AI Cat Narrator and the AI Cat Evaluator. We treated AI Cat Narrator as the generative model and developed the AI Cat Evaluator to mirror the conditions of the AI Cat Narrator, but with an evaluation function. The AI Cat Evaluator was trained to scrutinize the content generated from the AI Cat Narrator and assess its relevance to the original database content, providing its judgement along with a detailed reasoning for its judgement. The purpose of this AI evaluator was designed to help researchers optimize the descriptions generated by the AI Cat Narrator. By using a model of the same type for evaluation, we were able to more precisely identify and explain potential issues or errors in the narrative generation process, leading to more efficient model training. 
Moreover, some tips about training AI models suggest that AI models often perform better when instructing GPT to externalize its thinking process or prompting it with signals of urgency or importance, such as emotion prompts. Therefore, we integrated an emotion prompt, \textit{"This is very important to my career"} \cite{li2023large}, and a Zero-Shot Chain-of-Thought prompt, ``\textit{Let's think progress step by step}'', to increase the output quality \cite{li2023large}. The results showed that the model indeed generated content more aligned with researchers' expectations when these prompting techniques were used.

\subsection{Preliminary Training Results}
To accurately present our initial result in training personalized AI Cat Narrators, we examined the differences in narratives produced by two versions of AI Cat Narrators. The database of AI Cat Narrator (A) is comprised solely of factual data, while the database for AI Cat Narrator (B) has been augmented with pet phrases and narratives from novels. The analysis revealed that narratives from AI Cat Narrator (B), due to their incorporation of specific phrases, more effectively capture and express the unique personalities of cats. For instance, in Fig.\ref{fig:two}-1, phrases such as \textit{"Such a coward"}  and \textit{"Time to check my gear"} highlight the cats’ penchant for adventure.  Moreover, compared to AI Cat Narrator (A), which directly describes the scenes in images, AI Cat Narrator (B) enhances its storytelling by weaving in these pet phrases and novelistic elements, creating narratives with deeper emotional resonance and contextual richness. This approach not only deepens the narrative’s complexity but also more strongly engages the audience's interest.

\section{Discussion \& Future Work}
This study presents a method for developing an AI Cat Narrator capable of interpreting real-life data from a unique perspective. We analyzed video data collected from the viewpoint of cats to create a database for building the AI Cat Narrator. The collected data is categorized into two types: Factual data and Counterfactual data both based on individual interpretations of reality. We also integrated fictional plots from novels into the AI Cat Narrator's training database. The AI Cat Narrator's defamiliarized fictional narratives encourage participants to discover hints about their everyday interactions between humans and cats. These stories not only evoke memories and connections but also promote deep contemplation and discussion regarding the nature of reality. 
While attributing human characteristics to animals can stimulate our imagination and has influenced pop culture in various media forms, this practice can also lead to misunderstandings about animal behavior. However, we propose that anthropomorphism, when used to spark intellectual curiosity rather than debate the truth of fictional stories, can enhance our reasoning and interpretive abilities through analogy. For instance, scientist Seven Dijkgraaf~\cite{de2016we} gained insights into bat behavior by imagining what it would be like to experience life as a bat. As such, we argue that employing fictional techniques to study human-animal interactions, with appropriate frameworks, can offer designers and researchers more comprehensive perspectives when investigating the design of such interactions, especially those involving cats. For our future work, we are currently conducting extensive experiments and interviews to gain a deeper understanding of the dynamics between humans and cats in shared living spaces, as well as evaluate the utility of this tool. We are currently engaging participants to interpret videos using the AI Cat Narrator, aiming to draw design insights from these varied interpretations. This approach is intended to reveal design possibilities to designers and researchers, facilitating the development of practical solutions for future smart living spaces accommodating both humans and cats.




\section{Conclusion}
This paper presents an innovative tool called AI Cat Narrator, which offers a fresh interpretation of everyday life from the cats' perspective. Research findings suggest that integrating real-world data and creative fictional content to construct an AI Cat Narrator results in the development of an AI model with unique characteristics. Therefore, AI Cat Narrators can create unique stories that merge reality with fantasy, not only stimulating people's imagination but also potentially making the stories resonate with people.

\bibliographystyle{ACM-Reference-Format}
\bibliography{AI_cat_narrator}


\begin{thebibliography}{9}


\ifx \showCODEN    \undefined \def \showCODEN     #1{\unskip}     \fi
\ifx \showDOI      \undefined \def \showDOI       #1{#1}\fi
\ifx \showISBNx    \undefined \def \showISBNx     #1{\unskip}     \fi
\ifx \showISBNxiii \undefined \def \showISBNxiii  #1{\unskip}     \fi
\ifx \showISSN     \undefined \def \showISSN      #1{\unskip}     \fi
\ifx \showLCCN     \undefined \def \showLCCN      #1{\unskip}     \fi
\ifx \shownote     \undefined \def \shownote      #1{#1}          \fi
\ifx \showarticletitle \undefined \def \showarticletitle #1{#1}   \fi
\ifx \showURL      \undefined \def \showURL       {\relax}        \fi
\providecommand\bibfield[2]{#2}
\providecommand\bibinfo[2]{#2}
\providecommand\natexlab[1]{#1}
\providecommand\showeprint[2][]{arXiv:#2}

\bibitem[Bell et~al\mbox{.}(2005)]%
        {bell2005making}
\bibfield{author}{\bibinfo{person}{Genevieve Bell}, \bibinfo{person}{Mark Blythe}, {and} \bibinfo{person}{Phoebe Sengers}.} \bibinfo{year}{2005}\natexlab{}.
\newblock \showarticletitle{Making by making strange: Defamiliarization and the design of domestic technologies}.
\newblock \bibinfo{journal}{\emph{ACM Transactions on Computer-Human Interaction (TOCHI)}} \bibinfo{volume}{12}, \bibinfo{number}{2} (\bibinfo{year}{2005}), \bibinfo{pages}{149--173}.
\newblock


\bibitem[Chang et~al\mbox{.}(2017)]%
        {chang2017interview}
\bibfield{author}{\bibinfo{person}{Wen-Wei Chang}, \bibinfo{person}{Elisa Giaccardi}, \bibinfo{person}{Lin-Lin Chen}, {and} \bibinfo{person}{Rung-Huei Liang}.} \bibinfo{year}{2017}\natexlab{}.
\newblock \showarticletitle{" Interview with Things" A First-thing Perspective to Understand the Scooter's Everyday Socio-material Network in Taiwan}. In \bibinfo{booktitle}{\emph{Proceedings of the 2017 Conference on Designing Interactive Systems}}. \bibinfo{pages}{1001--1012}.
\newblock


\bibitem[Cheng et~al\mbox{.}(2019)]%
        {cheng2019peekaboo}
\bibfield{author}{\bibinfo{person}{Yu-Ting Cheng}, \bibinfo{person}{Mathias Funk}, \bibinfo{person}{Wenn-Chieh Tsai}, {and} \bibinfo{person}{Lin-Lin Chen}.} \bibinfo{year}{2019}\natexlab{}.
\newblock \showarticletitle{Peekaboo Cam: Designing an Observational Camera for Home Ecologies Concerning Privacy}. In \bibinfo{booktitle}{\emph{Proceedings of the 2019 on Designing Interactive Systems Conference}}. \bibinfo{pages}{823--836}.
\newblock


\bibitem[De~Waal(2016)]%
        {de2016we}
\bibfield{author}{\bibinfo{person}{Frans De~Waal}.} \bibinfo{year}{2016}\natexlab{}.
\newblock \bibinfo{booktitle}{\emph{Are we smart enough to know how smart animals are?}}
\newblock \bibinfo{publisher}{WW Norton \& Company}.
\newblock


\bibitem[Desjardins and Biggs(2021)]%
        {desjardins2021data}
\bibfield{author}{\bibinfo{person}{Audrey Desjardins} {and} \bibinfo{person}{Heidi~R Biggs}.} \bibinfo{year}{2021}\natexlab{}.
\newblock \showarticletitle{Data epics: Embarking on literary journeys of home internet of things data}. In \bibinfo{booktitle}{\emph{Proceedings of the 2021 CHI Conference on Human Factors in Computing Systems}}. \bibinfo{pages}{1--17}.
\newblock


\bibitem[Giaccardi et~al\mbox{.}(2016)]%
        {giaccardi2016thing}
\bibfield{author}{\bibinfo{person}{Elisa Giaccardi}, \bibinfo{person}{Nazli Cila}, \bibinfo{person}{Chris Speed}, {and} \bibinfo{person}{Melissa Caldwell}.} \bibinfo{year}{2016}\natexlab{}.
\newblock \showarticletitle{Thing ethnography: Doing design research with non-humans}. In \bibinfo{booktitle}{\emph{Proceedings of the 2016 ACM conference on designing interactive systems}}. \bibinfo{pages}{377--387}.
\newblock


\bibitem[H.~Tan et~al\mbox{.}(2022)]%
        {h2022critical}
\bibfield{author}{\bibinfo{person}{Neilly H.~Tan}, \bibinfo{person}{Brian Kinnee}, \bibinfo{person}{Dana Langseth}, \bibinfo{person}{Sean A.~Munson}, {and} \bibinfo{person}{Audrey Desjardins}.} \bibinfo{year}{2022}\natexlab{}.
\newblock \showarticletitle{Critical-playful speculations with cameras in the home}. In \bibinfo{booktitle}{\emph{Proceedings of the 2022 CHI Conference on Human Factors in Computing Systems}}. \bibinfo{pages}{1--22}.
\newblock


\bibitem[Li et~al\mbox{.}(2023)]%
        {li2023large}
\bibfield{author}{\bibinfo{person}{Cheng Li}, \bibinfo{person}{Jindong Wang}, \bibinfo{person}{Yixuan Zhang}, \bibinfo{person}{Kaijie Zhu}, \bibinfo{person}{Wenxin Hou}, \bibinfo{person}{Jianxun Lian}, \bibinfo{person}{Fang Luo}, \bibinfo{person}{Qiang Yang}, {and} \bibinfo{person}{Xing Xie}.} \bibinfo{year}{2023}\natexlab{}.
\newblock \showarticletitle{Large language models understand and can be enhanced by emotional stimuli}.
\newblock \bibinfo{journal}{\emph{arXiv preprint arXiv:2307.11760}} (\bibinfo{year}{2023}).
\newblock


\bibitem[Natsume(1906)]%
        {natsume1906cat}
\bibfield{author}{\bibinfo{person}{S{\=o}seki Natsume}.} \bibinfo{year}{1906}\natexlab{}.
\newblock \bibinfo{booktitle}{\emph{I am a cat}}. Vol.~\bibinfo{volume}{1}.
\newblock \bibinfo{publisher}{Hattori Shoten}.
\newblock


\end{thebibliography}

\appendix

\end{document}